\documentclass[aps,rmp]{revtex4}
\input epsf

\def\lap{\hbox{~{\lower -2.5pt\hbox{$<$}}\hskip -8pt\raise 
-3.5pt\hbox{$\sim$}}}
\def\gap{\hbox{~{\lower -2.5pt\hbox{$>$}}\hskip -8pt\raise 
-3.5pt\hbox{$\sim$}}}
\def\apg{\hbox{{\raise -2.5pt\hbox{$>$}}\hskip -8pt\lower - 
2.5pt\hbox{$\sim$}}}
\def\apl{\hbox{{\raise -2.5pt\hbox{$<$}}\hskip -8pt\lower - 
2.5pt\hbox{$\sim$}}}

\begin{document}

\bigskip

\title{MAXWELL'S DEMON, SZILARD'S ENGINE AND QUANTUM MEASUREMENTS
\footnote{
This paper has appeared in 1984 as a Los Alamos preprint LAUR 84-2751. It is based
on talks given by the author in 1983 and 1984, including an invited presentation
at the APS Spring 1983 meeting. It was circulated as a preprint in 1984, and
eventually published in {\it Frontiers of Nonequilibrium Statistical Physics},
NATO ASI Series B: Physics Vol. 135, G. T. Moore and M. O. Scully, eds. (Plenum,
New York, 1986). It was also reprinted in {\it Maxwell's Demon: Entropy,
Information, Computing}, H. S. Leff and A. F. Rex, eds. (Princeton University
Press, 1990). The posted version is essential identical to the original,
although a few typos have been corrected.}}

\author{Wojciech Hubert Zurek}
\affiliation{Theory Division, Mail Stop B213, LANL, Los Alamos, New Mexico 87545}

\bigskip
\begin{abstract}
We propose and analyze a quantum version of Szilard's ``one-molecule 
engine.'' In particular, we recover, in
the quantum context, Szilard's conclusion concerning the free energy 
``cost'' of measurements: $\Delta F \geq
k_B T\ln2$ per bit of information.
\end{abstract}
\maketitle

\section{INTRODUCTION} In 1929 Leo Szilard wrote a path-breaking 
paper$^1$ entitled ``On the Decrease of Entropy in
a Thermodynamic System by the Intervention of Intelligent 
Beings''. There, on the basis of a thermodynamic
``gedankenexperiment'' involving ``Maxwell's demon,'' he argued that 
an observer, in order to learn, through a
measurement, which of the two equally probable alternatives is 
realized, must use up at least

\begin{equation}  \Delta F = k_BT \ln2
\end{equation}
of free energy. Szilard's paper not only correctly defines the 
quantity known today as
information, which has found a wide use in the work of Claude Shannon 
and others in the field of communication
science.$^3$ It also formulates physical connection between 
thermodynamic entropy and information-theoretic
entropy by establishing ``Szilard's limit'', the least price which 
must be paid in terms of free energy for the
information gain.

The purpose of this paper is to translate Szilard's classical thought 
experiment into a quantum one, and to
explore its consequences for quantum theory of measurements. A 
``one-molecule gas'' is a ``microscopic'' system
and one may wonder whether conclusions of Szilard's classical 
analysis remain valid in the quantum domain. In
particular, one may argue, following Jauch and Baron,$^4$ that 
Szilard's analysis is inconsistent, because it employs two different, 
incompatible classical idealizations of the one-molecule gas -- 
dynamical and thermodynamical -- to arrive at Eq. (1). We
shall show that the apparent inconsistency pointed out by Jauch and 
Baron is removed by quantum treatment. This is not too surprising, 
for, after all, thermodynamic entropy which is central in this 
discussion is incompatible with classical mechanics, as it becomes 
infinite in the limit
$\hbar \rightarrow 0$. Indeed, information--theoretic analysis of the 
operation of Szilard's engine allows one to
understand, in a very natural way, his thermodynamical conclusion, 
Eq. (1), as a consequence of the
{\it conservation of information} in a closed quantum system.

The quantum version of Szilard's engine will be considered in the 
following section. Implications of Szilard's
reasoning for quantum theory and for thermodynamics will be explored 
in Sec. III -- where we shall assume that
``Maxwell's Demon'' is classical -- and in Sec. IV, where it will be a 
quantum system.

\section{QUANTUM VERSION OF SZILARD'S ENGINE}

A complete cycle of Szilard's classical engine is presented in Fig. 
1. The work it can perform in the course of
one cycle is
\begin{equation}
\Delta W = \int_{V/2}^{V} p(v) dv = k_BT \int_{V/2}^{V} dv/v = k_BT\ln2
\end{equation}
Above, we have used the law of Gay-Lussac, $p=kT/V$, for one-molecule 
gas. This work gain per cycle can be
maintained in spite of the fact that the whole system is at the same 
constant temperature $T$. If Szilard's model
engine could indeed generate useful work, in the way described above 
at no additional expense of free energy, it
would constitute a perpetuum mobile, as it delivers mechanical work 
from an (infinite) heat reservoir with no
apparent temperature difference. To fend off this threat to the 
thermodynamic irreversibility, Szilard has
noted that, ``If we do not wish to admit that the Second Law has been 
violated, we must conclude that
$\dots$ the measurement $\dots$ must be accompanied by a production 
of entropy.'' Szilard's conclusion has
far-reaching consequences, which have not yet been fully explored. If 
it is indeed correct, it can provide an
operational link between the concepts of ``entropy'' and 
``information.'' Moreover, it forces one to admit that a
measuring apparatus can be used to gain information only if 
measurements are essentially irreversible.

Before accepting Szilard's conclusion one must realize that it is 
based on a very idealized model. In particular,
two of the key issues have not been explored in the original paper. 
The first, obvious one concerns fluctuations.
One may argue that the one-molecule engine cannot be analyzed by 
means of thermodynamics, because it is nowhere
near the thermodynamic limit. This objection is overruled by noting 
that arbitrarily many ``Szilard's engines''
can be linked together to get a ``many-cylinder'' version of the 
original design. This will cut down fluctuations
and allow one to apply thermodynamic concepts without difficulty.

A more subtle objection against the one-molecule engine has been 
advanced by Jauch and Baron.$^4$ They note that
``Even the single-molecule gas is admissible as long as it satisfies 
the gas laws. However, at the exact moment
when the piston is in the middle of the cylinder and the opening is 
closed, the gas violates the law of
Gay-Lussac because gas is compressed to half its volume without 
expenditure of energy.'' Jauch and Baron
``$\dots$ therefore conclude that the idealizations in Szilard's 
experiment are inadmissible in their actual
context $\dots$'' This objection is not easy to refute for the 
classical one-molecule gas. Its molecule should
behave as a billiard ball. Therefore, it is difficult to maintain 
that after the piston has been inserted, the
gas molecule still occupies the whole volume of the container. More 
defensible would be a claim that while the
molecule is on a definite side of the partition, we do not know on 
which side, and this prevents extraction of
the promised $k_BT \ln2$ of energy. (This was more or less Szilard's 
position.) The suspicious feature of such
argument is its {\it subjectivity}: Our (classical) intuition tells 
us that the gas molecule is on a definite side
of a partition. Moreover, what we (or Maxwell's demons) know should 
have nothing to do with the objective state
of the gas. It is this {\it objective} state of the gas which should 
allow one to extract energy. And,
objectively, the gas has been compressed to half its volume by the 
insertion of the partition no matter on which
side of the piston the molecule is. The eventual ``observation'' may 
help in making the engine work, but one may
argue, as do Jauch and Baron, that the ``potential to do work'' seems 
to be present even before such a
measurement is performed.

To re-examine arguments of Jauch and Baron, consider the quantum 
version of Szilard's engine, shown in Fig. 1.
The role of the container is now played by the square potential well. 
A single gas molecule of mass $m$ is
described by the (nonrelativistic) Schrodinger equation with the 
boundary conditions $\psi(-L/2) =
\psi(L/2) = 0$. Energy levels and eigenfunctions are given by

\[
E =  n^2 \pi^2 \hbar^2/(2mL^2)  = \epsilon n^2  
\]

\[
\langle x | \psi_n \rangle = \left\{ 
\begin{array}{ll}
(2/L)^{1/2} \cos 2 \pi n x/L  & \mbox{for $n = 2k+1$} \\
(2/L)^{1/2} \sin 2 \pi n x/L  & \mbox{for $n = 2k$} 
\end{array} 
\right.
\]

At a finite temperature $T = 1/ (\beta~k_B)$ the equilibrium 
state of the system will be completely described
by the density matrix:
\begin{equation}
\rho = Z^{-1} \sum_n \exp(- \beta E_n) |\psi \rangle \langle \psi_n | \ \ .
\end{equation}
Above, $Z$ is the usual partition function:
\begin{equation}
Z = \sum_{n=1}^\infty \exp(- \beta E_n^2) = \sum_{n=1}^\infty \varsigma n^2
\end{equation}
For $1/2 < \varsigma < 1, Z$ can be adequately approximated by:
\begin{equation}
Z = \frac{1}{2} \left(\sqrt{\pi/|\ln \varsigma|} -1 \right) \ \ .
\end{equation}
For our purposes a still simpler, high-temperature approximation
\begin{equation}
Z = (\pi / \epsilon \beta)^{1/2} / 2 = L / \left(h^2/2 mk_BT\right)^{1/2} \ \ .
\end{equation}
valid for $\epsilon \ll k_B T$, will be sufficient most of the time. This 
is, of course, the familiar Boltzmann gas
partition function. It can be readily generalized to the three-dimensional box
\begin{equation}
Z = L_x L_y L_z/(h^2/2 \pi mk_BT)^{3/2} \ \ ,
\end{equation}
as well as to the case when there are $N$ ``classically'' 
indistinguishable particles. The point of this
elementary calculation is to demonstrate that, in the further 
analysis, we can rely on classical estimates of
pressure, internal energy, entropy, etc. $\dots$ for one molecule gas 
which were used by Szilard$^1$: A partition
function completely determines thermodynamic behavior of the system.

Consider now a ``piston,'' slowly inserted in the middle of the 
potential well ``box``. This can be done either
(1) while the engine is coupled with the reservoir, or (2) when it is 
isolated. In either case, it must be done
slowly, so that the process remains either (1) thermodynamically or 
(2) adiabatically reversible. We shall
imagine the piston as a potential barrier of width $d \ll L$ and 
height that eventually attains $U \gg k_BT$.
The presence of this barrier will alter the structure of the energy 
levels: these associated with even $n$ will
be shifted "upwards" so that the new eigenvalues are:
\begin{equation}
E^\prime_{2k} = \epsilon^\prime (2k)^2 + \Delta_k = E_k + \Delta_k
\end{equation}
where
\begin{equation}
\epsilon^\prime = \epsilon L^2 / (L - d)^2
\end{equation}
and
\begin{equation}
\Delta_k \cong (4 \epsilon^\prime/ \pi) \exp \left(-d\sqrt{2m(U - E_k)/
\hbar} \right) \ \ .
\end{equation}
Energy levels corresponding to the odd values of n are shifted upwards by
$\Delta E_n \sim (2n+l)\epsilon^\prime$ so that
\begin{equation}
E^\prime_{2k-1} = \epsilon^\prime (2k)^2 -\Delta_k = E_k -\Delta_k
\end{equation}
A pair of the eigenvalues $E^\prime_{2k}$, $E^\prime_{2k-1}$ can be 
alternatively regarded as the
k$^{th}$ doubly degenerate eigenstate of the newly created two-well potential
with degeneracy broken for finite values of $U$.

For $U \rightarrow \infty$ exact eigenfunctions can be given for 
each well separately.
For finite $U$, for these levels where $\Delta_k \ll E_k$, 
eigenfunctions of the complete potential can be
reconstructed from the k$^{th}$ eigenfunctions of the left 
$(|L_k\rangle)$ and right $(|R_k\rangle)$ wells:
$$
E_k +\Delta_k \leftrightarrow |\psi_k^+\rangle = (|L_k\rangle - |R_k 
\rangle)  / \sqrt{2} \eqno(12a)
$$
$$
E_k -\Delta_k \leftrightarrow |\psi_k^-\rangle = (|L_k\rangle + |R_k 
\rangle)  / \sqrt{2} \eqno(12b)
$$
Alternatively, eigenfunctions of the left and right wells can be expressed
in terms of energy eigenfunctions of the complete Hamiltonian
\renewcommand\theequation{12{\rm c}}
\begin{equation}
|L_k \rangle  =   (|\psi_k^+\rangle+ | \psi_k^- \rangle) \ \sqrt{2}
\end{equation}
\renewcommand\theequation{12{\rm d}}
\begin{equation}
|R_k \rangle  =   (|\psi_k^-
\rangle- | \psi_k^+ \rangle) \ \sqrt{2}
\end{equation}
\section{MEASUREMENTS BY THE CLASSICAL MAXWELL'S DEMON}

Consider a measuring apparatus which, when inserted into Szilard's 
engine, determines on which side of the
partition the molecule is. Formally, this can be accomplished by the 
measurement of the observable
\renewcommand\theequation{13}
\begin{equation}
\hat{\Pi} = \lambda (|L \rangle \langle L| - |R \rangle \langle R|) \ \ .
\end{equation}
Here $\lambda$ is an arbitrary eigenvalue while;
\renewcommand\theequation{13{\rm a}}
\begin{equation}
|L \rangle \langle L|= \sum_{k=1}^N |L_k \rangle \langle L_k| \ ,
\end{equation}
\renewcommand\theequation{13{\rm b}}
\begin{equation}
|R \rangle \langle R| = \sum_{k=1}^N |R_k \rangle \langle R_k| \ ,
\end{equation}
and $N$ is sufficiently large, $N^2 \epsilon \beta \gg 1$.
The density matrix before the measurement, but after piston is 
inserted, is given by
\renewcommand\theequation{14}
\begin{eqnarray}
&  \tilde{\rho} = \tilde{Z}^{-1} \sum^\infty_{k=1} \exp(\beta E_k) 
\left\{\exp(-\beta\Delta_k) |\psi_k^+ \rangle
\langle \psi_k^+| \right\}& \\
& = \tilde{Z}^{-1} \sum^\infty_{k=1}\exp(-\beta E_k)\left\{\cosh(\beta\Delta_k) 
(|L_k \rangle \langle L_k| + |R_k \rangle \langle R_k|) + \sinh(\beta\Delta_k)
(|L_k \rangle \langle R_k| + |R_k \rangle \langle L_k|) 
\right\}& \nonumber
\end{eqnarray}
Depending on the outcome of the observation, the density matrix 
becomes either $\rho_L$ or $\rho_L$ where;
\renewcommand\theequation{15{\rm a}}
\begin{equation}
\rho_L = Z_L^{-1}\sum_{k=1}^\infty \exp(-\beta E_k) \cosh 
\beta\Delta_k |L_k \rangle \langle L_k| \ \ ,
\end{equation}
\renewcommand\theequation{15{\rm b}}
\begin{equation}
\rho_R = Z_R^{-1}\sum_{k=1}^\infty \exp(-\beta E_k) \cosh 
\beta\Delta_k |R_k \rangle \langle R_k| \ \ .
\end{equation}
Both of these options are chosen with the same probability. The 
classical ``demon'' ``knows'' the outcome of the
measurement. This information can be used to extract energy in the 
way described by Szilard.

We are now in a position to return to the objection of Jauch and 
Baron. The ``potential to do work'' is measured
by the free energy $A(T,L)$ which can be readily calculated from the 
partition function
$$
A(T,L) = -k_B T \ln Z(T,L) \ \ .
$$
For the one-molecule engine this free energy is simply
\renewcommand\theequation{16}
\begin{equation}
A = -k_B T \ln [L/(h^2/2 \pi mk_B T)^{1/2}]
\end{equation}
{\underline {before}} the partition is inserted. It becomes
\renewcommand\theequation{17}
\begin{equation}
\tilde{A} = -k_B T \ln [(L-d)/(h^2/2 \pi mk_B T)^{1/2}]
\end{equation}
after the insertion of the partition. Finally, as a result of the 
measurement, free energy increases regardless of the outcome
\renewcommand\theequation{18}
\begin{eqnarray}
A_L & = & -k_B T \ln [((L-d)/2) h^2 /2\pi mk_B T)^{1/2}] \nonumber \\
& = & -k_BT(\tilde{A} - \ln 2)
\end{eqnarray}
Let us also note that the change of $A$ as a result of the insertion 
of the partition is negligible:
\renewcommand\theequation{19}
\begin{equation}
\tilde{A} - A  = k_B T \ln (L/(L-d)) \sim 0 (d/L) \ \ .
\end{equation}
However, the change of the free energy because of the measurement is 
precisely such as to account for the $W =
k_BT \ln 2$ during the subsequent expansion of the piston:
\renewcommand\theequation{20}
\begin{equation}
A_L - \tilde{A} = k_B T \ln 2 \ \ .
\end{equation}
It is now difficult to argue against the conclusion of Szilard. The 
formalism of quantum mechanics confirms the
key role of the measurement in the operation of the one-molecule 
engine. And the objection of Jauch and Baron, based on a classical 
intuition, is overruled.

The classical gas molecule, considered by Szilard, as well as by 
Jauch and Baron, may be on the unknown side of
the piston, but cannot be on ``both'' sides of the piston. Therefore, 
intuitive arguments concerning the potential to do useful work could 
not be unambiguously settled in the context of classical dynamics and
thermodynamics. Quantum molecule, on the other hand, can be on 
``both'' sides of the potential barrier, even if
its energy is far below the energy of the barrier top, and it will 
``collapse'' to one of the two potential wells
only if $\hat{\Pi}$ is ``measured.''

It is perhaps worth pointing out that the density matrix 
$\tilde{\rho}$, Eq. (15), has a form which
is consistent with the ``classical'' statement that ``a molecule is 
on a definite, but unknown side of the
piston'' almost equally well as with the statement that ``the 
molecule in in a thermal mixture of the energy
eigenstates of the two-well potential.'' This second statement is 
rigorously true in the limit of weak coupling
with the reservoir: gas is in contact with the heat bath and therefore 
is in thermal equilibrium. On the other
hand, the off-diagonal terms of the very same density matrix in the 
$|L_k \rangle, |R_k \rangle$ representation
are negligible $(\sim \sinh \beta \Delta_k)$. Therefore, one can 
almost equally well maintain that this density
matrix describes a molecule which is on an ``unknown, but definite'' 
side of the partition. The phase between
$|R_n \rangle$  and $|L_n \rangle$ is lost.

The above discussion leads us to conclude that the key element needed 
to extract useful work is the
{\underline {correlation}} between the state of the gas and the state 
of the demon. This point can be analyzed
further if we allow, in the next section, the ``demon'' to be a quantum system.

\section{MEASUREMENT BY "QUANTUM MAXWELL'S DEMON''}

  Analysis of the measurement process in the previous section was very 
schematic. Measuring apparatus, which has
played the role of the ``demon'' simply acquired the information 
about the location of the gas molecule, and this
was enough to describe the molecule by $\rho_L$ or $\rho_R$. The 
second law demanded the entropy of the
apparatus to increase in the process of measurement, but, in the 
absence of a more concrete model for the
apparatus it was hard to tell why this entropy increase was essential 
and how did it come about. The purpose of
this section is to introduce a more detailed model of the apparatus, 
which makes such an analysis possible.

We shall use as an apparatus a two-state quantum system. We assume 
that it is initially prepared by some external
agency -- we shall refer to it below as an ``external observer'' -- 
in the ``ready to measure'' state $|D_0
\rangle$. The first stage of the measurement will be accomplished by 
the transition:
\renewcommand\theequation{21{\rm a}}
\begin{equation}
|L_n \rangle| D_0 \rangle \rightarrow  |L_n \rangle| D_L \rangle
\end{equation}
\renewcommand\theequation{21{\rm b}}
\begin{equation}
|R_n \rangle| D_0 \rangle \rightarrow  |R_n \rangle| D_R \rangle
\end{equation}
for all levels $n$. Here, $|D_R \rangle$  and $|D_L \rangle$  must be 
orthogonal if the measurement
is to be reliable. This will be the case when the gas and the demon 
interact via an appropriate coupling
Hamiltonian, e.g.:
\renewcommand\theequation{22}
\begin{equation}
H_{int} = i \delta \left(|L_n \rangle \langle L_n| - |R_n \rangle 
\langle R_n| \right) \left(|D_L \rangle
\langle D_R| - |D_R \rangle \langle D_L| \right)
\end{equation}
for the time interval $\Delta t = \pi \hbar/ (4 \delta)^{5,6}$, and 
the initial state of the demon is:
\renewcommand\theequation{23}
\begin{equation}
|D_0 \rangle = \left(|D_L \rangle  + |R_R \rangle \right) / \sqrt{2}
\end{equation}
For, in this case the complete density matrix becomes:
\renewcommand\theequation{24}
\begin{equation}
P = \exp\left( -i H_{int} \Delta t / \hbar \right) \tilde{\rho} |D_0 
\rangle \langle D_0 | 
\cong \left(\rho_L |D_L \rangle \langle D_L| + \rho_R |D_R \rangle 
\langle D_R | \right) /2
\end{equation}
Here and below we have omitted small off-diagonal terms $\sim \beta 
\Delta_n$ present in
$\tilde{\rho}$, Eq. (14) and, by the same token, we shall drop 
corrections $(\sim \beta^2 \Delta^2_n)$ from the diagonal
of $\rho_L$ and $\rho_R$. As it was pointed out at the end of the 
last section, for the equilibrium there is no need for
the ``reduction of the state vector.'' The measured system -- one 
molecule gas -- is already in the mixture of being on
the left and right-hand side of the divided well.

To study the relation of the information gain and entropy increase in
course of the measurement, we employ the usual definition of entropy 
in terms of the density matrix$^{3,6,7,8}$
\renewcommand\theequation{25}
\begin{equation}
S(\rho) = -k_B Tr \rho \ln \rho
\end{equation}
The information is then:
\renewcommand\theequation{26}
\begin{equation}
I(\rho) =  \ln (Dim{\mathcal H}) - S(\rho) /k_B  \ \ ,
\end{equation}
where ${\mathcal H}$ is the Hilbert space of the system in question. 
Essential in our further discussion will be the
mutual information $I_\mu(P_{AB})$ defined for two subsystems, $A$ 
and $B$, which are described jointly by the density
matrix $P_{AB}$, while their individual density matrices are $\rho_A$ 
and $\rho_A$:
\renewcommand\theequation{27}
\begin{equation}
I_\mu (P_{AB}) =  I(P_{AB}) - (I(\rho_A) + I(\rho_B)) \ \ .
\end{equation}
In words, mutual information is the difference between the 
information needed to specify $A$ and $B$ one at a time and jointly.
When they are not correlated, $P_{AB} = \rho_A \rho_B$, 
then $I_\mu(P_{AB}) = 0$.
The readoff of the location of the gas molecule by the demon will be 
described by an external observer not aware of the
outcome by the transition:
\renewcommand\theequation{28}
\begin{equation}
\tilde{\rho} |D_0 \rangle \langle D_0| \rightarrow \left( \rho_L |D_L 
\rangle  \langle D_L| + \rho_R |D_R \rangle
\langle D_R | \right)/2
\end{equation}
The density matrix of the gas is then:
\renewcommand\theequation{29}
\begin{equation}
\tilde{\rho} = \tilde{Z}^{-1} \sum_{n=1} \exp \left(-\beta \epsilon 
n^2 \right) \left(|L_n \rangle \langle L_n| + |R_n
\rangle \langle R_n| \right) = \left(\rho_L + \rho_R \right)/2
\end{equation}
Thus, even after the measurement by the demon, the density matrix of the gas,
$\rho_G$ will be still:
\renewcommand\theequation{30}
\begin{equation}
\rho_G = \langle D_L
|P| D_L \rangle + \langle D_R |P| D_R \rangle \cong \tilde{\rho}
\end{equation}
The state of the demon will, on the other hand, change from the 
initial pure state $|D_0 \rangle \langle D_0|$ into a
mixture:
\renewcommand\theequation{31}
\begin{equation}
\rho_D = \sum_n \left(\langle L_n |P| L_n \rangle + \langle R_n |P| 
R_n \rangle \right) = \left(|D_L \rangle \langle
D_L | + |D_R \rangle \langle D_R| \right)/2
\end{equation}
Entropy of the gas viewed by the external observer remains constant:
\renewcommand\theequation{32}
\begin{equation}
S(\rho_G) = S(\tilde{\rho}) = \partial (\beta\ln\tilde{Z})/\partial \beta
\end{equation}
Entropy of the demon has however {\underline{increased}}:
\renewcommand\theequation{33}
\begin{equation}
S(\rho_D) - S(|D_0 \rangle \langle D_0|) = k_B \ln 2
\end{equation}
Nevertheless, the
combined entropy of the gas-demon system could not have changed: In 
our model evolution during the read-off was
dynamical, and the gas-demon system was isolated. Yet, the sum of the 
entropies of the two subsystems -- the gas and the
demon -- has increased by $k_B \ln 2$. The obvious question is then: 
where is the ``lost information''
$\Delta I = I(P) - (I(\tilde{\rho}) + I(\rho_D))$? The very form of 
this expression and its similarity to the right-hand
side of Eq. (27) suggests the answer: The loss of the information by 
the demon is compensated for by an equal increase
of the mutual information:
\renewcommand\theequation{34}
\begin{equation}
\Delta I_\mu = I_\mu(P) - I_\mu (\tilde{\rho}|D_0 \rangle \langle D_0|) \ \ .
\end{equation}
Mutual information can be regarded as the information gained by
the demon. From the conservation of entropy during dynamical 
evolutions it follows that an increase of the mutual
information must be compensated for by an equal increase of the entropy.
\renewcommand\theequation{35}
\begin{equation}
\Delta I_\mu - \Delta S/k_B = 0
\end{equation}
This last equation is the
basis of Szilard's formula, Eq. (1).

At this stage readoff of the location of the gas molecule is 
reversible. To undo
it, one can apply the inverse of the unitary operation which has 
produced the correlation. This would allow one to
erase the increase of entropy of the demon, but only at a price of 
the loss of mutual information.

One can now easily
picture further stages of the operation of Szilard's engine. The 
state of the demon can be used to decide which well is
``empty'' and can be ``discarded.''$^6$  The molecule in the other 
well contains twice as much energy as it would if the
well were {\underline{adiabatically}} (i.e. after decoupling it from 
the heat reservoir) expanded from its present size
$\sim L/2$ to $L$. Adiabatic expansion conserves entropy. Therefore, 
one can immediately gain $\Delta W
= \langle E \rangle / 2 = k_B T/4$ without any additional entropy 
increases. One can also gain $\Delta W
= k_B T \ln 2$,
Eq. (2), by allowing adiabatic expansion to occur in small 
increments, in between which one molecule gas is reheated
by the re-established contact with the reservoir. In the limit of 
infinitesimally small increments this leads, of
course, to the isothermal expansion. At the end the state of the gas 
is precisely the same as it was at the beginning of
the cycle, and we are $\Delta W$ ``richer'' in energy. If Szilard's 
engine could repeat this cycle, this would be a
violation of the second law of thermodynamics. We would have a 
working model of a {\underline {perpetuum mobile}},
which extracts energy in a cyclic process from an infinite reservoir 
{\underline {without}} the temperature difference.

Fortunately for the second law, there is an essential gap in our 
above design: The demon is still in the mixed state,
$\rho_D$ Eq. (31). To perform the next cycle, the observer must 
``reset'' the state of the demon to the initial
$|D_0\rangle$. This means that the entropy $dS = k_B \ln 2$ must be 
somehow removed from the system. If we were to leave
demon in the mixed state, coupling it with the gas through $H_{int}$ 
will not result in the increase of their mutual
information: The one-bit memory of the demon is still filled by the 
outcome of the past measurement of the gas.
Moreover, this measurement cannot any longer be reversed by allowing 
the gas and the demon to interact. For, even
though the density matrix of each of these two systems has remained 
the same, their joint density matrix is now very
different:
\renewcommand\theequation{36}
\begin{equation}
\tilde{\rho} \otimes \rho_D = \left\{(\rho_L |D_L \rangle \langle D_L| + 
\rho_R |D_R \rangle \langle D_R|)/2 + (\rho_L |D_R
\rangle \langle D_R| + \rho_R |D_L \rangle \langle D_L|)/2 \right\}
\end{equation}
One could, presumably, still accomplish the reversal
using the work gained in course of the cycle. This would, however, 
defy the purpose of the engine. The only other way to
get the demon into the working order must be executed by an 
``external observer'' or by its automated equivalent: The
demon must be reset to the ``ready-to-measure'' state $|D_0 \rangle$.

As it was pointed out by Bennett in his classical
analysis, this operation of memory erasure is the true reason for 
irreversibility, and the ultimate reason why the free
energy, Eq. (1), must be expended.$^9$ Short of reversing the cycle, 
any procedure which resets the state of the demon to
$|D_0 \rangle$ must involve a measurement. For instance, a possible 
algorithm could begin with the measurement whether
the demon is in the state $|D_R \rangle$ or in the orthogonal state 
$|D_L \rangle$. Once this is established then,
depending on the outcome, the demon maybe rotated either 
``rightward'' from $|D_L \rangle$, or ``leftward'' from $|D_R
\rangle$ so that its state returns to $|D_0 \rangle$. The measurement 
by some external agency -- some ``environment''
-- is an essential and unavoidable ingredient of any resetting 
algorithm. In its course entropy is ``passed on'' from
the demon to the environment.

\section{SUMMARY}

  The purpose of our discussion was to analyze a process which converts
thermodynamic energy into useful work. Our analysis was quantum, 
which removed ambiguities pointed out by Jauch and
Baron in the original discussion by Szilard. We have concluded that 
validity of the second law of thermodynamics can
be satisfied only if the process of measurement is accompanied by the 
increase of entropy of the measuring apparatus by
the amount no less than the amount of gained information. Our 
analysis confirms therefore conclusions of Szilard (and
earlier discussion of Smoluchowski$^{10}$, which has inspired 
Szilard's paper). Moreover, we show that the ultimate
reason for the entropy increase can be traced back to the necessity 
to ``reset'' the state of the measuring apparatus$^9$ ,
which, in turn, must involve a measurement. This necessity of the 
``readoff'' of the outcome by some external agency --
the environment of the measuring apparatus -- has been already invoked 
to settle some of the problems of quantum theory
of measurements$^{11}$. Its role in the context of the above example 
is to determine what was the outcome of the
measurement, so that the apparatus can be reset. In the context of 
quantum measurements its role is very similar: it
forces one of the outcomes of the measurement to be definite, and, 
therefore, causes the ``collapse of the
wavepacket.''

In view of the title of Szilard's paper, it is perhaps worthwhile to 
point out that we did not have to
invoke ``intelligence'' at any stage of our analysis: all the systems 
could be presumably inanimate. However, one can
imagine that the ability to measure, and make the environment ``pay'' 
the entropic cost of the measurement is also one
of the essential attributes of animated beings.

\section*{ACKNOWLEDGMENTS}

  I would like to thank Charles Bennett and John Archibald
Wheeler for many enjoyable and stimulating discussions on the subject of 
this paper. The research was supported in part by
the National Science Foundation under Grant No. PHY77-27084, 
supplemented by funds from the National Aeronautics and
Space Administration.

\newpage

\begin{figure}
\epsfxsize=7cm
\epsfbox{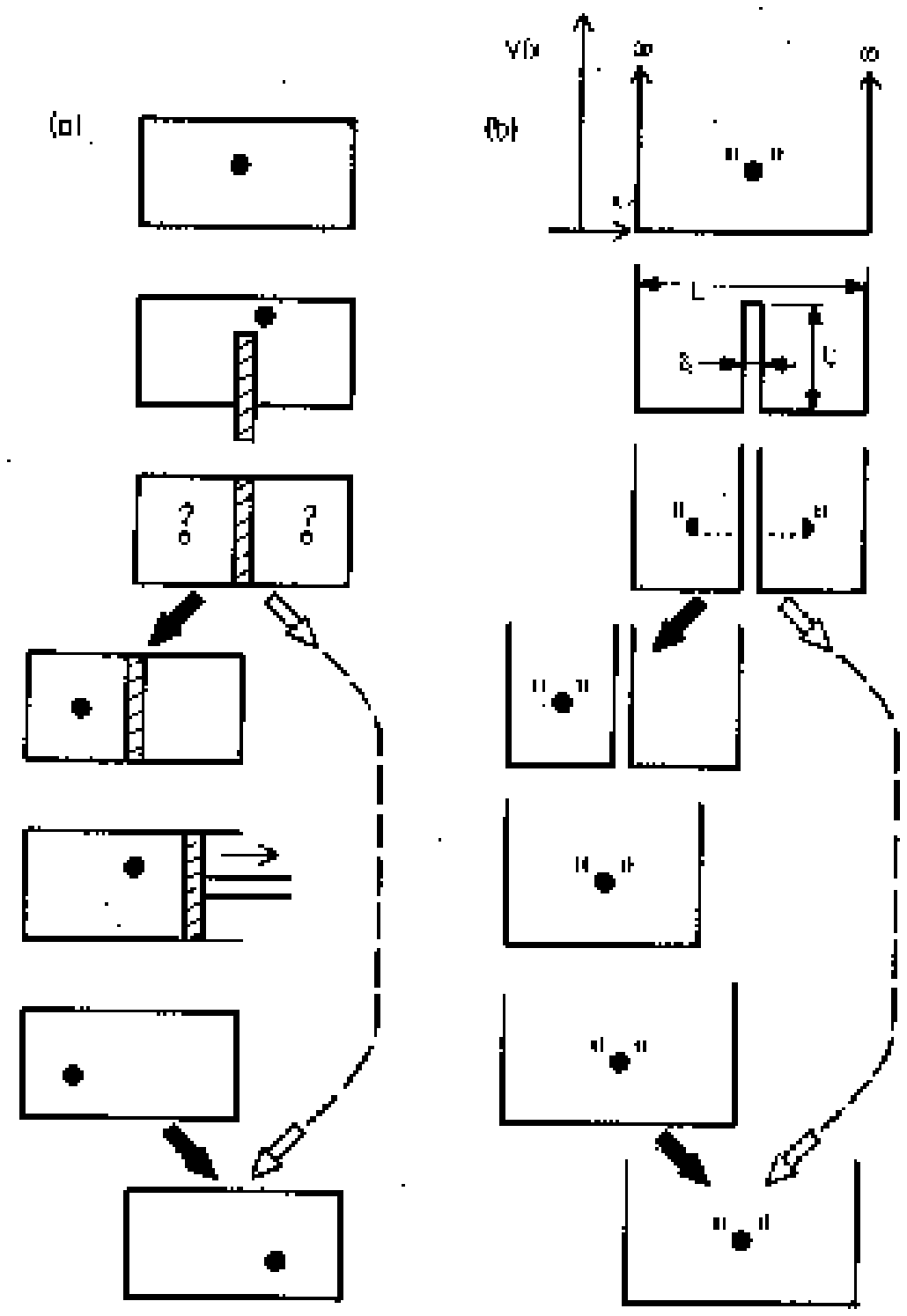} 
\caption{A cycle of Szilard's engine. a) Original, classical version; 
b) Quantum version discussed here.  The measurement of the 
location of the molecule is essential in the process of extracting 
work in both classical and quantum design.}
\end{figure}


\section*{REFERENCES}
\begin{enumerate}
\bibitem{1} L. Szilard, ``On the Decrease of Entropy in a 
Thermodynamic System by the Intervention of Intelligent Beings,'' 
{\it Z. Phys.} 53:840 (1929): English translation reprinted 
in {\underline{Behavioral Science}} 9:301 (1964),
as well as in Ref. 2, p. 539.

\bibitem{2} J.A. Wheeler and W.H. Zurek, eds., {\underline{Quantum 
Theory and Measurement}} (Princeton University Press,
Princeton, 1983).

\bibitem{3} C.E. Shannon and W. Weaver, {\underline{The Mathematical 
Theory of Communication}}, (University of Illinois
Press, Urbana, 1949).

\bibitem {4}  J.M. Jauch and J.G. Baron, ``Entropy, Information, and 
Szilard's Paradox,'' {\it Helv. Phys. Acta}, 45:220
(1972).

\bibitem {5} W.H. Zurek, ``Pointer Basis of Quantum Apparatus: Into 
What Mixture Does the Wavepacket Collapse?''
{\it Phys. Rev. D} 24:1516 (1981).

\bibitem {6} W.H. Zurek, ``Information Transfer in Quantum 
Measurements: Irreversibility and Amplification'', in 
{\underline{Quantum Optics, Experimental Gravitation 
and Measurement Theory}}, eds: P. Meystre and
M.O. Scully (Plenum Press, New York, 1983). {\tt quant-ph 0111137}

\bibitem {7} J.R. Pierce and E.C. Rosner, {\underline{Introduction to 
Communication Science and Systems}} (Plenum Press,
New York, 1980).

\bibitem {8} H. Everett III, {\underline {Dissertation}}, 
Princeton University, 1957, 
reprinted in B.S. DeWitt and N. Graham, eds., {\underline
{The Many -- Worlds Interpretation of Quantum Mechanics}} 
(Princeton University Press, Princeton, 1973).

\bibitem {9} C.H. Bennett, ``The Thermodynamics of
Computation'' {\it Int. J. Theor. Phys.} 21:305 (1982).

\bibitem {10} M. Smoluchowski, ``Vortr\"{a}ge liber die Kinetische Theorie der
Materie und Elektrizitat,'' Leipzig 1914.

\bibitem {11} W.H. Zurek, ``Environment-Induced Superselection 
Rules'' {\it Phys. Rev.} D26:1862 (1982).
\end{enumerate}
\end{document}